\documentclass[doublecol]{epl2}

\title{Exact solution of gyration radius of individual's trajectory\\ for a simplified human mobility model}
\shorttitle{Exact solution of gyration radius for a human mobility model} 

\author{Xiao-Yong Yan\inst{1,2} \and Xiao-Pu Han\inst{2} \and Tao Zhou\inst{2,3} \and Bing-Hong Wang\inst{2}}

\shortauthor{Xiao-Yong Yan \etal}

\institute{
  \inst{1} Department of Transportation Engineering, Shijiazhuang Tiedao University - Shijiazhuang 050043, China \\
  \inst{2} Department of Modern Physics, University of Science and Technology of China - Hefei 230026, China \\
  \inst{3} Web Sciences Center, University of Electronic Science and Technology of China - Chengdu 610054, China \\

}
\pacs{89.75.Da}{Systems obeying scaling laws}
\pacs{05.40.Fb}{Random walks and Levy flights}
\pacs{89.40.Bb}{Land transportation}

\abstract{
Gyration radius of individual's trajectory plays a key role in quantifying human mobility patterns. Of particular interests, empirical analyses suggest that the growth of gyration radius is slow versus time except the very early stage and may eventually arrive to a steady value. However, up to now, the underlying mechanism leading to such a possibly steady value has not been well understood. In this Letter, we propose a simplified human mobility model to simulate individual's daily travel with three sequential activities: commuting to workplace, going to do leisure activities and returning home. With the assumption that individual has constant travel speed and inferior limit of time at home and work, we prove that the daily moving area of an individual is an ellipse, and finally get an exact solution of the gyration radius. The analytical solution well captures the empirical observation reported in [M. C. Gonz\'alez \emph{et al.}, \emph{Nature}, \textbf{453} (2008) 779]. We also find that, in spite of the heterogeneous displacement distribution in the population level, individuals in our model have characteristic displacements, indicating a completely different mechanism to the one proposed by Song \emph{et al}. [\emph{Nat. Phys.} \textbf{6} (2010) 818].
}
\begin{document}
\maketitle

Understanding human mobility patterns is important to many different disciplines from epidemiology to communication engineering and transportation engineering~\cite{rev,disease,wireless,trans}. With the development of modern electronic techniques, some novel devices including online bill tracker~\cite{nat06}, mobile phone~\cite{nat08,jsm10} and Global Positioning System~\cite{jiang,levy} have been used to track human mobility trajectories. By analyzing these data, scientists have revealed some remarkable features of human mobility patterns, such as heavy-tail distributions of travel distances and waiting times, spatio-temporal regularity and bounded nature of individual trajectory~\cite{nat06,nat08,jsm10,jiang,levy,jpa,sci10}. To explain the origin of observed scaling properties in human mobility patterns, formulation of simplified models and their quantitative analysis have attracted increasing interests from many branches of sciences~\cite{np10,hxp1,hxp2,hyq,slaw,hcmm}.

A key quantity related to human mobility trajectories is gyration radius~\cite{rg}, a measure of how far from the center of mass the mass is. Gonz\'alez, Hidalgo and Barab\'asi~\cite{nat08} used gyration radius to quantify individual trajectory tracked from numerous anonymized mobile phone users. The gyration radius grows fast in the very early stage and then become very slow versus time. It approximately approaches to a steady value in hundreds of hours. They suggested a logarithmical function to fit the empirical data. Moreover, they found that gyration radius has a strong impact on travel distance distributions over all users. After the removal of the dependence of gyration radius, the travel distance distributions of groups collapsed into a single curve. Therefore, they suggested using gyration radius as a characteristic travel distance for each individual. In a word, these empirical evidences indicated that the gyration radius of individual's trajectory plays a key role in characterizing human mobility patterns.

Despite its importance for quantitatively understanding human mobility patterns, the factors responsible for the gyration radius of trajectory is still unclear. Several models~\cite{hxp1,hxp2,hyq,slaw,hcmm} have been developed to reproduce the spatio-temporal patterns of human mobility, but none of them explained the mechanism stabilizing gyration radius. Recently, Song \emph{et al.}~\cite{np10} proposed a novel model for individual human mobility. Their model is based on two competitive mechanisms: exploring unknown locations and returning preferentially to familiar places. The model explained many scaling laws observed in human mobility patterns, as well as the ultraslow increasing phenomenon of gyration radius. However, the factors affecting the gyration radius of individual's trajectory have not been fully documented.

In this Letter, we proposed a simple model for human mobility as well as a method to solve it. This model is motivated by empirical evidence drawn from a travel diary data set named \emph{Mobidrive}~\cite{ax1}. The data set records the day-to-day travel behaviors of 360 persons in two Germany cities over a six-week survey period. A series of researches based on this data set have revealed some common features of individuals' daily travel behaviors~\cite{ax2,ax3,ax4,ax5}. The most notable one of these features is the similarity of individual's weekday travel activities~\cite{ax2,ax3}, which means that most individuals take their working-day activities in nearly the same order: commuting to workplace (or school for students) in the morning, spending their daytime at workplace, doing some leisure activities (shopping, dining, entertainment, \emph{etc}.) and returning home in the evening. Although the times and visiting locations of leisure activities can be variable in this ``home-work-leisure-home" trips chain~\cite{ax4}, the activities that periodic commuting to workplace and returning home are nearly invariant. Similar regular patterns were also observed in some mobile phone users' travel behaviors~\cite{rt}.

Here we introduce a straightforward model to reproduce the regular travel behavior of human individual in working days. For simplicity, we assume that: (i) each individual has fixed locations of home and workplace; (ii) each individual has an inferior limit of time at home and work everyday; (iii) each individual has a constant travel speed relied on her/his common means of transportation; (iv) each individual takes one trip per day(A trip specifically stands for the travel from workplace to a selected location for leisure activities. The other two movements, from home to workplace and from the selected location to home, are not accounted). The last assumption is based on the result of a previous empirical research on \emph{Mobidrive}~\cite{ax4}, which indicated that in most time an individual performs no more than one trip per day. Based on the assumptions, we present a detailed description of our model: in every working day, an individual first commutes to workplace from home, then goes to a random location for leisure activities, and finally returns home. The individual can select any locations to do after-work activities, but she/he must guarantee to return home in time. In other words, because of the restrictions of travel speed and the maximum possible leisure time everyday, the individual is prohibited to select the points far away from home and workplace in our model. Then, after $T$ simulation steps, we will obtain a trajectory composed with $T$ points at ``home", $T$ points at ``workplace" and $T$ points scattered in a defined area.

\begin{figure}
\onefigure[scale=2.0]{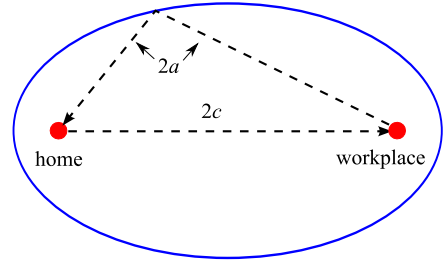}
\caption{(Color online) Elliptical area of an individual's daily travel. Two foci of the ellipse are home and workplace, and the distance between them are focal length ($2c$). The major axis length ($2a$) represents the maximum daily travel distance.}
\label{fig.1}
\end{figure}

The defined area of individual's trajectory in our model can be shaped by a geometric approach: with the model rules described above, the maximum possible travel distance in one's everyday after-work activities (including home-return) is
\begin{equation}
\label{eq-1}
2a=v(24-t_{m})-2c,
\end{equation}
where $v$ is the travel speed, $t_{m}$ is individual's inferior limit of time at home and work everyday, $c$ is half of the distance between home and work ($c < a$). The time resolution is hour. Obviously, we can draw the maximum area of individual daily travel as an ellipse (see Fig.~\ref{fig.1}). The ellipse's two foci are individual's home and workplace, the focal length is the distance between home and workplace, and the major axis length, $2a$, is the maximum possible daily travel distance subtracting commute distance. The elliptical area of individual human mobility has been empirically found in \emph{Mobidrive}~\cite{ax5} (see also Fig.3a in Ref.~\cite{nat06} where the mobile phone users' trajectories seem to be also limited in elliptical areas).

With the obtained elliptical area, we next solve the gyration radius of trajectory for our model. The gyration radius of the trajectory can be expressed as
\begin{equation}
\label{eq-2}
r_{g}=\sqrt{\frac{1}{3T}\sum_{i=1}^{3T}(\vec{r}_{i}-\vec{r}_{cm})^{2}},
\end{equation}
where $T$ is the simulation step, $\vec{r}_{cm}$ is the center of mass of the trajectory, and in each time step, three locations are considered, namely $\vec{r}_{1}=\vec{r}_{4}=\vec{r}_{7}=\cdots=$ home, $\vec{r}_{2}=\vec{r}_{5}=\vec{r}_{8}=\cdots=$ workplace, and $\vec{r}_{3},\vec{r}_{6},\vec{r}_{9},\cdots,\vec{r}_{3T}$ are the locations randomly selected in the ellipse. When $T$ is large enough, according to the geometrical symmetry, we can deduce that the center of mass of the locations is at the center of the ellipse. Notice that the $2T$ points at foci (home and work) of the ellipse have same distance, $c$, to the center of the ellipse, we can rewrite Eq.~(\ref{eq-2}) to
\begin{equation}
\label{eq-3}
r_{g}=\sqrt{\frac{2}{3}c^{2}+\frac{1}{3}\overline{I}},
\end{equation}
where $\overline{I}$ is the mean square distance from $T$ random points to the center of the ellipse. This distance can be calculated as
\begin{equation}
\label{eq-4}
\overline{I}=\frac{\int_{0}^{2\pi} \textrm{d}\theta \int_{0}^{1}  (a^{2} \textrm{cos}^{2}\theta +b^{2} \textrm {sin}^{2} \theta) ab  \rho^{3}\textrm{d}\rho }{\pi ab}
=\frac{a^{2}+b^{2}}{4},
\end{equation}
where $b$ is the half minor axis length of the ellipse. Combining Eqs.~(\ref{eq-2}-\ref{eq-4}) and $b^{2}=a^{2}-c^{2}$, we finally obtain
\begin{equation}
\label{eq-5}
r_{g}=\sqrt{\frac{7c^{2}+2a^{2}}{12}}.
\end{equation}

The result shows that the gyration radius of individual's trajectory relates to two factors: one is the distance between individual's home and workplace, another is individual's maximum daily travel distance, which depends on individual's travel speed and daily time assignment. The result also implies that the individual who has more leisure time and more advanced means of transportation will has a larger area of daily mobility.

\begin{figure}
\onefigure[width=0.45\textwidth]{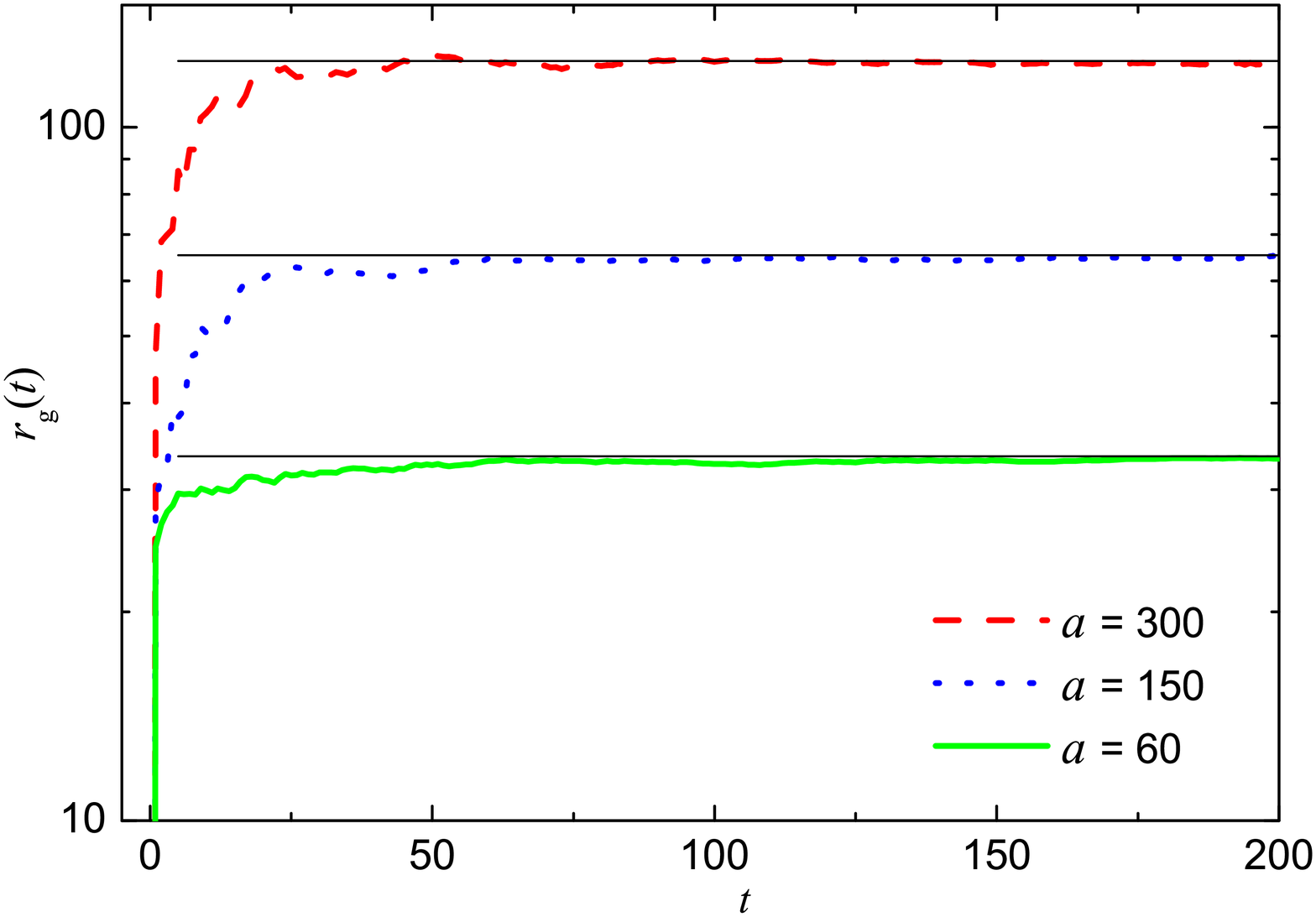}
\caption{(Color online) The time evolution of gyration radius in the model ($c=30$). The curves are numerical simulations for different $a$, and the straight lines represent analytical results. Data points are averaged by 100 independent runs, each lasts 200 steps after initialization.}
\label{fig.2}
\end{figure}

Notice that our analytical result shows that the radius of gyration is a constant, but the previous empirical observation~\cite{nat08} shown that it grows with time in the very early stage. To explain this difference, we simulate the time evolution of gyration radius for our model (see Fig.~\ref{fig.2}) . The simulation results demonstrate that the gyration radii grow very fast in the early stage and will converge asymptotically to stable values in accordance with the corresponding solutions. Although one may choose logarithmic function to fit the simulation curves before they converged, the curves are indeed in the relaxed stage. The empirical observations reported in Ref.~\cite{nat08} look like the simulations shown in Fig.~\ref{fig.2}. Although these empirical observations can be well captured by logarithmic functions, they may converge to a steady value after a very long time, as suggested by the present model. In addition, it is more convenient to use a single value rather than the whole of the time evolution curve of gyration radius to characterize human mobility patterns.
\begin{figure}
\onefigure[width=0.45\textwidth]{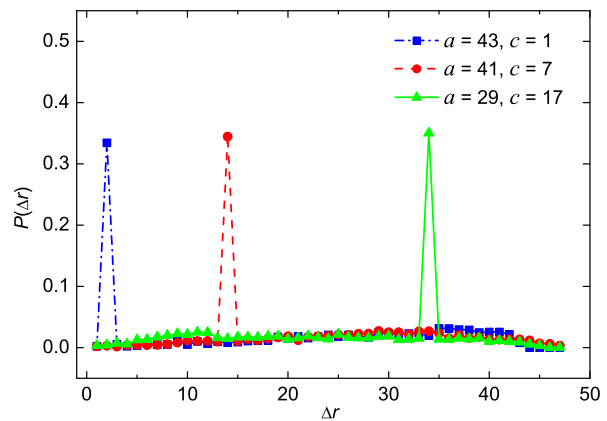}
\caption{(Color online) Travel distance distributions of three individuals with the same $r_{g}=15.12$ in the model. Data points are averaged by 100 dependent runs, each run lasts 1000 steps after initialization. }
\label{fig.3}
\end{figure}

Additionally, from Eq.~(\ref{eq-5}) we know that individuals sharing same $r_{g}$ can have different $c$. The parameter $c$ have crucial influence on travel distance distribution of individual because there are one third of trips taken between home and workplace in our model. To demonstrate this, we simulate the travel distance distribution of individuals with same $r_{g}$ but different $c$, and plot the results in Fig.~\ref{fig.3}. As the figure shows, all of the distributions are different: they peak respectively at $2c$, and, none of them is right-skewed. This seems incompatible with the empirical observation that human have heavy-tail distributions of travel distances~\cite{nat06,nat08}. Nevertheless, to the best of our knowledge, there were not sufficient evidences showing that the travel distance distribution of individual was heavy-tailed. In spite of the empirical evidence~\cite{nat08} shown that the population with same $r_{g}$ has heavy-tailed distribution of travel distances, it is insufficient to deduce the existence of a similar scaling law for individual. In the consideration of only daily travel instead of long-term traveling by trains, airplane, etc, $\cdots$, pealed law at individual level was also be observed \cite{bart}. Actually, the emergence of heavy-tailed nature at the population level can also be resulted from the heterogeneity of the distance between home and workplace (such heterogeneity has been reported in Ref.~\cite{sv1,sv2}).

\begin{figure}
\onefigure[width=0.45\textwidth]{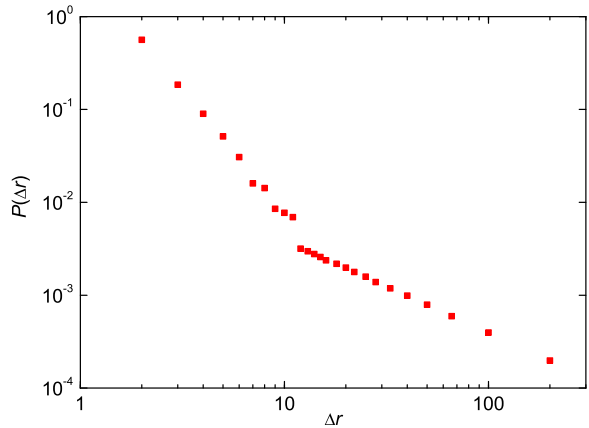}
\caption{(Color online) Travel distance distribution of population with the same $r_{g}=200$ but different $c$ following $P(c)\sim c^{-1}$ in the model. Data points are averaged by 100 independent runs, each lasts 1000 steps after initialization.}
\label{fig.4}
\end{figure}

Indeed, the empirical analyses at the individual level \cite{ax5} suggested that most individuals' movements are limited in an ellipse-like area with a peaked displacement distribution. Figure 4 reports an example where the distributions of $c$ follows $P(c)\sim c^{-1}$, where the displacement distribution can be approximately fitted with double power-law form. The result indicates that the heterogeneous displacement distribution emerged at the population level can be resulted from the heterogeneity of home-workplace distances or some other factors instead of the heterogeneous displacement distribution at the individual level.



In conclusion, we have proposed a simplified human mobility model and given an exact solution of gyration radius for it. The analytical solution agrees well with the empirical observation of gyration radius of mobile phone users, suggesting that our model may correctly capture the main mechanism leading to the bounded nature of human mobility. Different from the random diffusion of particles, human daily travel has a high degree of periodicity and regularity, which is the root of the high predictability of individual movements~\cite{sci10} and the rhythms of urban traffic~\cite{trans}. Additionally, individuals in urban travel are purposeful, or socially contextualized, and they have different destination choice strategies corresponding to various trips purposes. Our model, although simplified, incorporates these intrinsic periodicity and purposiveness of human travel behavior, and is expected to be extended to a more realistic traffic prediction approach in future research.

\acknowledgments
This work was partially supported by the National Natural Science Foundation of China (91024026, 10975126, 70871082, 70971089), the Specialized Research Fund for the Doctoral Program of Higher Education of China (20093402110032) and the Research Foundation of Hebei Educational Committee (Z2009139). XYY acknowledges the visiting scholar fellowship of the Chinese Ministry of Education.

\end{document}